\documentclass [12pt]  {revtex4}
\usepackage{CJK}
\usepackage{graphicx,amsmath}
\usepackage{booktabs}
\usepackage{amssymb}

\begin{document}
\begin{CJK*}{GBK}{song}
\CJKtilde\CJKindent
\title{Generation of atomic NOON states via adiabatic passage*}
\author{Qi-Gong Liu, Qi-Cheng Wu, Xin Ji\footnote{E-mail: jixin@ybu.edu.cn}, and Shou Zhang}
\affiliation{Department of Physics, College of Science, Yanbian
University, Yanji, Jilin 133002, People's Republic of China}
\begin{abstract}
 We propose a scheme for generating
atomic NOON states via adiabatic passage. In the scheme, a double
$\Lambda$-type three-level atom is trapped in a bimodal cavity and
two sets of $\Lambda$-type three-level atoms are translated into
and outside of two single mode cavities respectively. The three
cavities connected by optical fibres are always in vacuum states.
After a series of operations and suitable interaction time, we can
obtain arbitrary large-$n$ NOON states of two sets of
$\Lambda$-type three-level atoms in distant cavities by performing
a single projective measurement on the double $\Lambda$-type
three-level atom. Due to adiabatic elimination of atomic excited
states and the application of adiabatic passage, our scheme is
robust against the spontaneous emissions of atoms, the decays of
fibres and cavities photon leakage. So the scheme has a high
fidelity and feasibility under the current available techniques.
\end{abstract}

\maketitle
\section*{1. Introduction}

Quantum entanglement, an interesting and attractive phenomenon in
quantum mechanics, plays a significant role not only in testing
quantum nonlocality, but also in processing a variety of quantum
information tasks \cite{AKE,CHB,CHBS,CHBG,KMH,MHVB,SBZGCG,GV}.
Multi-particle entangled states, such as GHZ states, W states,
cluster states, NOON states, etc, are the fundamental resource of
quantum information processing (QIP). The NOON states, as an
intereting multi-particle entangled states, have the form as
\begin{equation}\label{01}
|{\rm{NOON}}\rangle=\frac{1}{\sqrt{2}}(|n,0\rangle+|0,n\rangle),
\end{equation}
which contain $n$ indistinguishable particles in an equal
superposition of all being in one of the two possible modes. It's
well established that the NOON states have significant
applications either in lithography \cite{ANB2000,MDA2001,KE2002}
or in quantum metrology \cite{MMW2004,RTG2008}. Especially, it can
be used to obviously improve the phase sensitivity in quantum
interferometry and beat the classical diffraction limit in quantum
lithography \cite{JJB1996,PW2004,JJA2009,RTG2008}. Recently, much
attention has been paid to prepare the NOON states. Many
theoretical and experimental schemes have been proposed for
generating the NOON states via optical components
\cite{CKH1987,CCG2001,KE2002,HL2002,KP2002,MMW2004,HHF2007,HC2007,AI2010},
cold atomic ensemble \cite{YAC2010}, superconducting circuits
\cite{STM2010,HW2011} and cavity quantum electrodynamics (QED)
\cite{KTK2007,RI2007,ZZR2010,RMD2011,LX2011,XXQ2012,NG2012,YRC2011,KL2013}.
However, the success probabilities of the proposals based on the
linear optical components, are very low  as the number of
particles increasing \cite{KP2002}. Some people suggested using
the optical nonlinear process for the generation of NOON states.
Yet, it is difficult to carry out experimentally so far
\cite{KTK2007}. The QED system, as a suitable candidate for
demonstrating QIP and quantum state engineering \cite{SB1999}, has
been studied extensively. Many proposals for preparing NOON states
based on QED have been put forward, as previously mentioned.
However, all of these systems are always affected by the various
of external factors which will induce the decoherence and affect
the probability of success, even destroy the entanglement.
Although many kinds of physical systems are used to avoid or
reduce the decoherence, it is always imperfect yet. So the
preparation of the entangled multi-particle NOON states is still a
severe challenge in the state of the art, though great progress
has been made in recent decades. On the other hand, the adiabatic
passage \cite{ASP1993,JRK1989,KB1998,NVV1999}, as an useful tool
for realizing QIP, is becoming more and more powerful and charming
owing to its robust against the spontaneous emission of the
excited states, cavity photon leakage and some experimental
parameter errors. As a result, the technique of adiabatic passage
has been extensively studied for the entanglement generation
\cite{MAT2005,WL2000,RGU2001,CM2003,LBC2007,SYY2008,SBZ2009,XYL2009,YLL2010,YLLF2010,PJS2011}.

In this paper, we propose a scheme for generating the NOON states
of two sets of the atoms via adiabatic passage. In the scheme, a
double $\Lambda$-type three-level atom is trapped in a bimodal
cavity and two sets of $\Lambda$-type three-level atoms are
translated into and outside of two single mode cavities
respectively. After a series of operations and suitable
interaction time, the arbitrary large-$n$ entangled NOON states of
two sets of $\Lambda$-type three-level atoms in distant cavities
can be obtained by performing a single projective measurement on
the double $\Lambda$-type three-level atom. Our scheme has the
following characteristics: (1) Due to along the dark state, the
cavity modes are unpopulated during the whole interaction process,
hence the scheme is robust against the decays of the cavities. (2)
Owing to adiabatical eliminations of atomic excited states, the
spontaneous emission rate can be regarded as zero. (3) Under
certain conditions, the probabilities of fibre modes populated can
be negligible safely, thus the decays of fibres are effectively
suppressed. (4) Taking advantages of adiabatic passage, the scheme
is insensitive to small fluctuation of experimental parameters.
(5) The scheme can be used to generate arbitrary large-$n$ NOON
states in theory.

The rest of the paper is organized as follows. In Sect. 2, the
fundamental model and Hamiltonian are introduced. In Sect. 3, we
propose a scheme to generate atomic NOON states via adiabatic
passage. Finally, we discuss the fidelity of the scheme and
summarize the conclusion in Sect. 4.

\section*{2. The fundamental model and Hamiltonian}

\begin{figure}
\scalebox{1.5}{\includegraphics{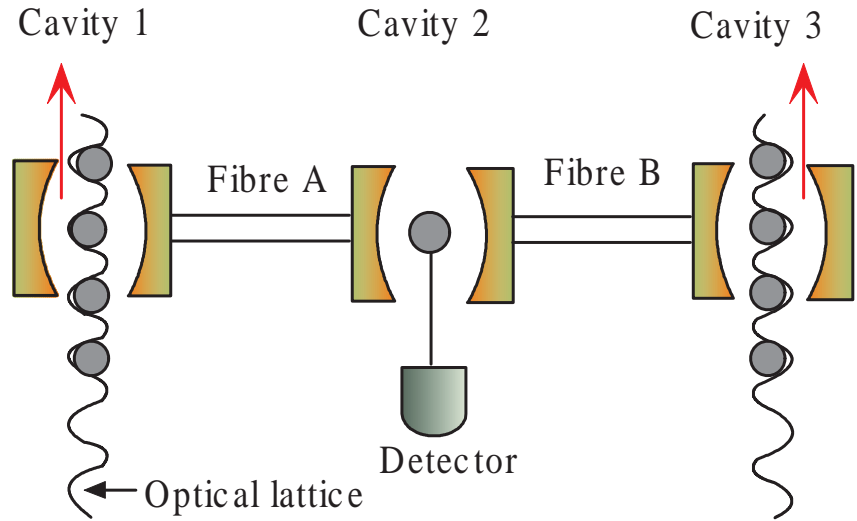}}\caption{The schematic
setup for generating atomic NOON states. The two sets of
$\Lambda$-type three-level atoms are stored in two transverse
optical lattices respectively and translated into and outside of
the cavity 1 and 3 respectively for interacting with the cavity
modes and the classical fields. The level configurations of the
atoms are shown in Fig.~\ref{fig02}. A double $\Lambda$-type
three-level atom with level configuration shown in
Fig.~\ref{fig03} is trapped in cavity 2. The three cavities are
connected by two fibres A and B. The atom detector is used to
detect the double $\Lambda$-type three-level atom.}\label{fig01}
\end{figure}

\begin{figure}
\scalebox{1.8}{\includegraphics{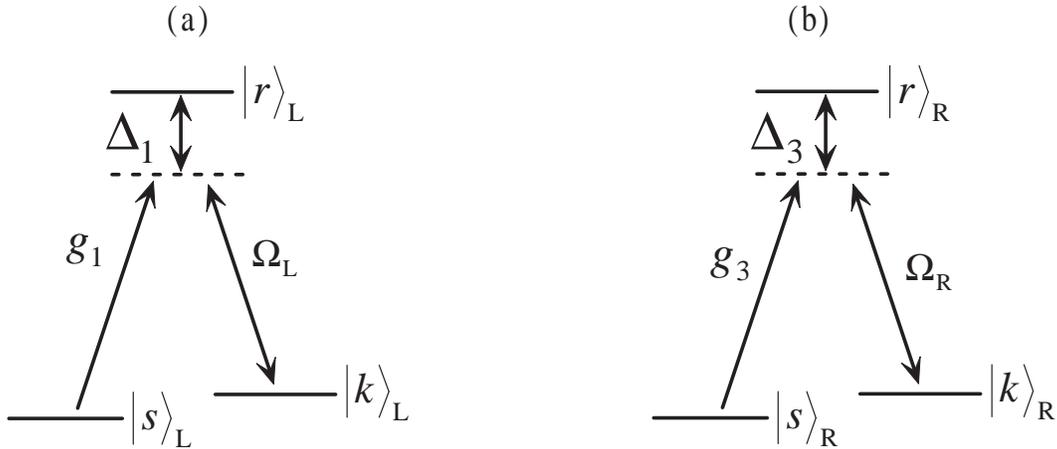}}\caption{The level
configurations of the $\Lambda$-type three-level atoms translated
into and outside of cavities 1 and 3. The states
$|s\rangle_{L(R)}$ and $|k\rangle_{L(R)}$ are two ground levels
and $|r\rangle_{L(R)}$ is an excited level of the atoms
interacting with cavity 1 (3). $g_{1(3)}$ is the coupling constant
between the transition
$|s\rangle_{L(R)}\leftrightarrow|r\rangle_{L(R)}$ and
corresponding cavity mode. $\Omega_{L(R)}$ is the time-dependent
Rabi frequency of classical field driving the atomic transition
$|k\rangle_{L(R)}\leftrightarrow|r\rangle_{L(R)}$. The classical
field with Rabi frequency $\Omega_{L(R)}$ and the mode of cavity
1(3) are detuned from the respective transitions by
$\Delta_{1(3)}$.}\label{fig02}
\end{figure}

\begin{figure}
\scalebox{1.8}{\includegraphics{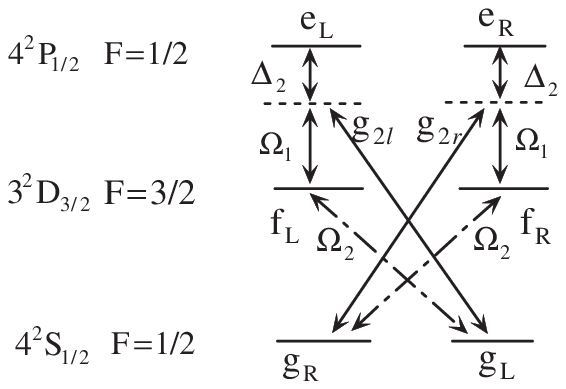}}\caption{The level
configuration of the double $\Lambda$-type three-level atom
\cite{KL2013}. $|g_{L}\rangle$ and $|g_{R}\rangle$,
$|e_{L}\rangle$ and $|e_{R}\rangle$ are two degenerate ground
states, two degenerate excited states respectively and
$|f_{L}\rangle$, $|f_{R}\rangle$ are two intermediate semi-stable
states. $\Omega_{1}$ is the Rabi frequency of classical field
$F_{1}$ driving the transitions
$|f_{L}\rangle\leftrightarrow|e_{L}\rangle$ and
$|f_{R}\rangle\leftrightarrow|e_{R}\rangle$; $g_{2l(r)}$ is the
coupling constant between the transition
$|e_{L}\rangle\leftrightarrow|g_{L}\rangle$
($|e_{R}\rangle\leftrightarrow|g_{R}\rangle$) and the left (right)
circular polarized cavity mode; $\Omega_{2}$ is the Rabi frequency
of classical field $F_{2}$ driving the transitions
$|g_{L}\rangle\leftrightarrow|f_{L}\rangle$ and
$|g_{R}\rangle\leftrightarrow|f_{R}\rangle$. The frequency
detunings of the cavity modes and classical cavity $F_{1}$ from
the respective atomic transitions are the same and denoted as
$\Delta_{2}$.}\label{fig03}
\end{figure}

The schematic setup for generating atomic NOON states is shown in
Fig.~\ref{fig01}. There are three distant optical cavities
connected by two fibres (fibres A and B). The two sets of
$\Lambda$-type three-level atoms are stored in two transverse
optical lattices respectively and translated into and outside of
the cavity 1 and cavity 3 simultaneously and respectively
\cite{PX2006,JAS2004}. The level configurations of the atoms
interacting with cavities 1 and 3 respectively are shown in
Fig.~\ref{fig02}(a) and Fig.~\ref{fig02}(b). The states
$|s\rangle_{L(R)}$ and $|k\rangle_{L(R)}$ are two ground levels
and $|r\rangle_{L(R)}$ is an excited level of the atoms
interacting with cavity 1 (3). The transition
$|s\rangle_{L(R)}\leftrightarrow|r\rangle_{L(R)}$ is coupled to
the mode of cavity 1 (3) with the coupling constant $g_{1(3)}$.
The transition $|k\rangle_{L(R)}\leftrightarrow|r\rangle_{L(R)}$
is driven by the classical field with time-dependent Rabi
frequency $\Omega_{L(R)}$. The frequency detunings between the
atomic transitions
$|s\rangle_{L(R)}\leftrightarrow|r\rangle_{L(R)},
|k\rangle_{L(R)}\leftrightarrow|r\rangle_{L(R)}$ and the relevant
cavity mode and classical cavity are the same and denoted as
$\Delta_{1(3)}$. They satisfy the corresponding two-photon
resonance conditions. The atom in the bimodal cavity 2 is a double
$\Lambda$-type three-level atom. The relevant atomic levels and
transitions are depicted in Fig.~\ref{fig03}. Such level structure
can be achieved in $^{40}Ca^{+}$
\cite{SBZ2006,HZW2007,MC2009,KL2013}. Two degenerate ground states
$|g_{L}\rangle$ and $|g_{R}\rangle$ correspond to $^{40}Ca^{+}$
atom hyperfine levels $|F=1/2,m=-1/2\rangle$ and
$|F=1/2,m=1/2\rangle$ of the level $4S_{1/2}$, while two
degenerate excited states $|e_{L}\rangle$ and $|e_{R}\rangle$
correspond to $|F=1/2,m=-1/2\rangle$ and $|F=1/2,m=1/2\rangle$ of
the level $4P_{1/2}$. On the other hand, two intermediate
semi-stable states $|f_{L}\rangle$ and $|f_{R}\rangle$ correspond
to $|F=3/2,m=-1/2\rangle$ and $|F=3/2,m=1/2\rangle$ of the level
$3D_{3/2}$, respectively. The transition
$|f_{L(R)}\rangle\leftrightarrow|e_{L(R)}\rangle$ is driven by a
classical field $F_{1}$ with Rabi frequency $\Omega_{1}$;
$|e_{L(R)}\rangle\leftrightarrow|g_{L(R)}\rangle$ is coupled to
the left(right) circular polarized cavity mode with the coupling
constant $g_{2l(r)}$; the transition
$|g_{L(R)}\rangle\leftrightarrow|f_{L(R)}\rangle$ is driven by
another classical field $F_{2}$ with Rabi frequency $\Omega_{2}$.
The frequency detunings of the cavity modes and classical cavity
$F_{1}$ from the respective atomic transitions are the same and
denoted as $\Delta_{2}$. Now, we consider the case both cavity 1
and cavity 3 have one atom respectively, thus the Hamiltonian of
atom-cavity system under the rotating-wave approximation can be
written as ($\hbar=1$)
\begin{eqnarray}\label{2}
H_{ac}&=&\frac{\Omega_{L}^2}{\Delta_{1}}|k\rangle_{L}\langle{k}|+\frac{\Omega_{1}^2}{\Delta_{2}}|f_{L}\rangle\langle{f_{L}}|+\frac{\Omega_{1}^2}{\Delta_{2}}|f_{R}\rangle\langle{f_{R}}|
+\frac{\Omega_{R}^2}{\Delta_{3}}|k\rangle_{R}\langle{k}|\cr
&&+\frac{g^2}{\Delta_{1}}a_{1}^{\dag}a_{1}|s\rangle_{L}\langle{s}|
+\frac{g^2}{\Delta_{2}}a_{2l}^{\dag}a_{2l}|g_{L}\rangle\langle{g_{L}}|
+\frac{g^2}{\Delta_{2}}a_{2r}^{\dag}a_{2r}|g_{R}\rangle\langle{g_{R}}|+\frac{g^2}{\Delta_{3}}a_{3}^{\dag}a_{3}|s\rangle_{R}\langle{s}|\cr
&&+\bigg(\frac{g\Omega_{L}}{\Delta_{1}}a_{1}^{\dag}|s\rangle_{L}\langle{k}|
+\frac{g\Omega_{1}}{\Delta_{2}}a_{2l}^{\dag}|g_{L}\rangle\langle{f_{L}}|
+\frac{g\Omega_{1}}{\Delta_{2}}a_{2r}^{\dag}|g_{R}\rangle\langle{f_{R}}|+\frac{g\Omega_{R}}{\Delta_{3}}a_{3}^{\dag}|s\rangle_{R}\langle
k|+\rm H.c.\bigg),
\end{eqnarray}
where $a_{1(3)}^{\dag}$ and $a_{1(3)}$ are the creation and
annihilation operators of the cavity 1(3); $a_{2l(r)}^{\dag}$ and
$a_{2l(r)}$ are the creation and annihilation operators of
left(right) circular polarization of the cavity 2. For
convenience, here we have set $g_{2l(r)}=g_{1(3)}=g$. The first
four terms on the right-hand side in Eq. (\ref{2}) represent the
atom level shifts induced by classical fields. By using the
nonresonant coupling of other lasers with the corresponding atom
levels, these energy level shifts can be compensated
straightforwardly \cite{TP1997}. Considering the cavities are
initially in vacuum states, the Hamiltonian $H_{ac}$ can be
further simplified into
\begin{eqnarray}\label{03}
H_{ac}^{'}=\Omega_{Le}(t)a_{1}^{\dag}|s\rangle_{L}\langle{k}|+\Omega_{1e}(t)a_{2l}^{\dag}|g_{L}\rangle\langle{f_{L}}|
+\Omega_{1e}(t)a_{2r}^{\dag}|g_{R}\rangle\langle{f_{R}}|+\Omega_{Re}(t)a_{3}^{\dag}|s\rangle_{R}\langle{k}|+\rm
H.c.,
\end{eqnarray}
where $\Omega_{Le}(t)=\Omega_{L}g/\Delta,
\Omega_{Re}(t)=\Omega_{R}g/\Delta,
\Omega_{1e}(t)=\Omega_{1}g/\Delta$ are the effective Rabi
frequencies for the corresponding Raman transitions
$|k\rangle_{L}\rightarrow|s\rangle_{L},
|k\rangle_{R}\rightarrow|s\rangle_{R},|f_{L(R)}\rangle\rightarrow|g_{L(R)}\rangle$,
respectively. Here we have assumed $\Delta_{1,2,3}=\Delta$.

In our scheme, three cavities are connected by two optical fibres
A and B. In the short fibre limit, only one (resonant) mode of a
fibre interacts with corresponding cavity modes. The interaction
Hamiltonian of fibre-cavity system can be approximated to
\begin{eqnarray}\label{04}
H_{cf}=\eta_{A}[b_{A}(a_{1}^{\dag}+a_{2l}^{\dag})]+\eta_{B}[b_{B}(a_{2r}^{\dag}+a_{3}^{\dag})]+\rm
H.c..
\end{eqnarray}
where $b_{A}$ and $b_{B}$ are the annihilation operators of the
resonant modes of fibre A and B respectively and the polarizations
of the fibre modes A and B have been chosen as the left circular
and the right circular polarizations, respectively. $\eta_{A(B)}$
is the coupling strength between the fibre mode A(B) and
corresponding cavity modes.

Lastly, in the interaction picture, the total Hamiltonian of the
system can be written as
\begin{eqnarray}\label{05}
H_{eff}&=&H_{ac}^{'}+H_{cf}\cr
&=&\Omega_{Le}(t)a_{1}^{\dag}|s\rangle_{L}\langle{k}|+\Omega_{1e}(t)a_{2l}^{\dag}|g_{L}\rangle\langle{f_{L}}|
+\Omega_{1e}(t)a_{2r}^{\dag}|g_{R}\rangle\langle{f_{R}}|+\Omega_{Re}(t)a_{3}^{\dag}|s\rangle_{R}\langle{k}|\cr
&&+\eta_{A}[b_{A}(a_{1}^{\dag}+a_{2l}^{\dag})]+\eta_{B}[b_{B}(a_{2r}^{\dag}+a_{3}^{\dag})]+\rm
H.c..
\end{eqnarray}

\section*{3. Generation of the NOON states}

In this section, we will show how to deterministically prepare the
multi-particle NOON states. Considering that both the cavity modes
and the fiber modes are all in vacuum states, initially, and the
double $\Lambda$-type three-level atom is prepared in the
superposition state $1/\sqrt{2}(|f_{L}\rangle+|f_{R}\rangle)$ with
the method of Ref.~\cite{CKL1998}, while atoms interacting with
cavities 1 and 3 respectively are in the state
$|s\rangle_{L}|s\rangle_{R}$, so the initial state of the whole
compound system is
\begin{eqnarray}\label{06}
\frac{1}{\sqrt{2}}(|f_{L}\rangle+|f_{R}\rangle)|s\rangle_{L}|s\rangle_{R}|000\rangle_{c}|00\rangle_{f}.
\end{eqnarray}

For the initial state
$|f_{L}\rangle|s\rangle_{L}|s\rangle_{R}|000\rangle_{c}|00\rangle_{f}$,
dominated by Hamiltonian (\ref{05}), the evolution of the system
state remains in the subspace with one excitation number spanned
by the basis vectors
\begin{eqnarray}\label{7}
|\psi_{1}\rangle&=&|f_{L}\rangle|s\rangle_{L}|s\rangle_{R}|000\rangle_{c}|00\rangle_{f},\cr
|\psi_{2}\rangle&=&|g_{L}\rangle|s\rangle_{L}|s\rangle_{R}|01^{l}0\rangle_{c}|00\rangle_{f},\cr
|\psi_{3}\rangle&=&|g_{L}\rangle|s\rangle_{L}|s\rangle_{R}|000\rangle_{c}|1^{l}0\rangle_{f},\cr
|\psi_{4}\rangle&=&|g_{L}\rangle|s\rangle_{L}|s\rangle_{R}|1^{l}00\rangle_{c}|00\rangle_{f},\cr
|\psi_{5}\rangle&=&|g_{L}\rangle|k\rangle_{L}|s\rangle_{R}|000\rangle_{c}|00\rangle_{f}.
\end{eqnarray}
The Hamiltonian (\ref{05}) has a dark state (i.e., zero energy
eigenstate of Hamiltonian (\ref{05}))
\begin{eqnarray}\label{8}
|\Psi_{D1}\rangle=\frac{1}{K_{0}}(\Omega_{Le}\eta_{A}|\psi_{1}\rangle-\Omega_{Le}\Omega_{1e}|\psi_{3}\rangle+\Omega_{1e}\eta_{A}|\psi_{5}\rangle),
\end{eqnarray}
where
$K_{0}=\sqrt{\Omega_{Le}^{2}\eta_{A}^{2}+\Omega_{Le}^{2}\Omega_{1e}^{2}+\Omega_{1e}^{2}\eta_{A}^{2}}$.

Under the condition of
\begin{eqnarray}\label{9}
\eta_{A}\gg \Omega_{Le},\Omega_{1e},
\end{eqnarray}
the dark state (\ref{8}) reduces to
\begin{eqnarray}\label{10}
|\Psi_{D1}^{'}\rangle=\frac{1}{\sqrt{\Omega_{Le}^{2}+\Omega_{1e}^{2}}}(\Omega_{Le}|\psi_{1}\rangle+\Omega_{1e}|\psi_{5}\rangle).
\end{eqnarray}
If pulse shapes are designed such that
\begin{eqnarray}\label{11}
\lim_{t\to-\infty}\frac{\Omega_{1e}}{\Omega_{Le}}=0,~
\lim_{t\to+\infty}\frac{\Omega_{Le}}{\Omega_{1e}}=0,
\end{eqnarray}
the initial state $|\psi_{1}\rangle$ of the system is
adiabatically transferred to $|\psi_{5}\rangle$.

For the initial state
$|f_{R}\rangle|s\rangle_{L}|s\rangle_{R}|000\rangle_{c}|00\rangle_{f}$,
dominated by Hamiltonian (\ref{05}) also, the evolution of the
system state remains in the subspace with one excitation number
spanned by the basis vectors
\begin{eqnarray}\label{12}
|\psi_{6}\rangle&=&|f_{R}\rangle|s\rangle_{L}|s\rangle_{R}|000\rangle_{c}|00\rangle_{f},\cr
|\psi_{7}\rangle&=&|g_{R}\rangle|s\rangle_{L}|s\rangle_{R}|01^{r}0\rangle_{c}|00\rangle_{f},\cr
|\psi_{8}\rangle&=&|g_{R}\rangle|s\rangle_{L}|s\rangle_{R}|000\rangle_{c}|01^{r}\rangle_{f},\cr
|\psi_{9}\rangle&=&|g_{R}\rangle|s\rangle_{L}|s\rangle_{R}|001^{r}\rangle_{c}|00\rangle_{f},\cr
|\psi_{10}\rangle&=&|g_{R}\rangle|s\rangle_{L}|k\rangle_{R}|000\rangle_{c}|00\rangle_{f}.
\end{eqnarray}
The Hamiltonian (\ref{05}) has a dark state
\begin{eqnarray}\label{13}
|\Psi_{D2}\rangle=\frac{1}{K_{1}}(\Omega_{Re}\eta_{B}|\psi_{6}\rangle-\Omega_{Re}\Omega_{1e}|\psi_{8}\rangle+\Omega_{1e}\eta_{B}|\psi_{10}\rangle),
\end{eqnarray}
where
$K_{1}=\sqrt{\Omega_{Re}^{2}\eta_{B}^{2}+\Omega_{Re}^{2}\Omega_{1e}^{2}+\Omega_{1e}^{2}\eta_{B}^{2}}$.

Under the condition of
\begin{eqnarray}\label{14}
\eta_{B}\gg \Omega_{Re},\Omega_{1e},
\end{eqnarray}
the dark state (\ref{13}) reduces to
\begin{eqnarray}\label{15}
|\Psi_{D2}^{'}\rangle=\frac{1}{\sqrt{\Omega_{Re}^{2}+\Omega_{1e}^{2}}}(\Omega_{Re}|\psi_{6}\rangle+\Omega_{1e}|\psi_{10}\rangle).
\end{eqnarray}
If pulse shapes are designed such that
\begin{eqnarray}\label{16}
\lim_{t\to-\infty}\frac{\Omega_{1e}}{\Omega_{Re}}=0,~
\lim_{t\to+\infty}\frac{\Omega_{Re}}{\Omega_{1e}}=0,
\end{eqnarray}
the initial state $|\psi_{6}\rangle$ of the system is
adiabatically transferred to $|\psi_{10}\rangle$.

After the above adiabatic processes, the system state
\begin{eqnarray}\label{17}
|\Psi_{1}\rangle&=&\frac{1}{\sqrt{2}}(|g_{L}\rangle|k\rangle_{L}|s\rangle_{R}+|g_{R}\rangle|s\rangle_{L}|k\rangle_{R})|000\rangle_{c}|00\rangle_{f}\cr
&=&\frac{1}{\sqrt{2}}(|g_{L}\rangle|1,0\rangle+|g_{R}\rangle|0,1\rangle)|000\rangle_{c}|00\rangle_{f}
\end{eqnarray}
can be achieved. Here, $|1,0\rangle (|0,1\rangle)$ denotes
$|k\rangle_{L}|s\rangle_{R} (|s\rangle_{L}|k\rangle_{R})$. After
that, we turn off the classical field $F_{1}$ and apply the
classical field $F_{2}$ with Rabi frequency $\Omega_{2}$ on the
double $\Lambda$-type three-level atom to drive the transition
$|g_{L}\rangle\rightarrow |f_{L}\rangle(|g_{R}\rangle\rightarrow
|f_{R}\rangle)$. After the interaction time $\tau$ which satisfies
$\Omega_{2}\tau=\pi/2$, the state of the whole system becomes
\begin{eqnarray}\label{18}
\frac{1}{\sqrt{2}}(|f_{L}\rangle|k\rangle_{L1}|s\rangle_{R1}+|f_{R}\rangle|s\rangle_{L1}|k\rangle_{R1})|000\rangle_{c}|00\rangle_{f}.
\end{eqnarray}
Here, in order to describe the process for preparing the atomic
NOON states clearly, we introduce symbol $|x\rangle_{L(R)i}
(x=s,k; i=1,2,3,\cdots)$ which denotes the $i$-th atom through the
cavity 1(3) is in the state $|x\rangle$. The state in Eq.
(\ref{18}) is the new initial state again. Then turn off the
classical field $F_{2}$ and turn on the classical field $F_{1}$
simultaneously. After repeating the above adiabatic process and
choosing suitable interaction time each time, the whole system
evolves successively into the states
\begin{eqnarray}\label{19}
|\Psi_{2}\rangle&=&\frac{1}{\sqrt{2}}(|g_{L}\rangle|k\rangle_{L1}|k\rangle_{L2}|s\rangle_{R1}|s\rangle_{R2}+|g_{R}\rangle|s\rangle_{L1}|s\rangle_{L2}|k\rangle_{R1}|k\rangle_{R2})|000\rangle_{c}|00\rangle_{f}\cr
&=&\frac{1}{\sqrt{2}}(|g_{L}\rangle|2,0\rangle+|g_{R}\rangle|0,2\rangle)|000\rangle_{c}|00\rangle_{f},\cr
|\Psi_{3}\rangle&=&\frac{1}{\sqrt{2}}(|g_{L}\rangle|k\rangle_{L1}|k\rangle_{L2}|k\rangle_{L3}|s\rangle_{R1}|s\rangle_{R2}|s\rangle_{R3}+|g_{R}\rangle|s\rangle_{L1}|s\rangle_{L2}|s\rangle_{L3}|k\rangle_{R1}|k\rangle_{R2}|k\rangle_{R3})|000\rangle_{c}|00\rangle_{f},\cr
&=&\frac{1}{\sqrt{2}}(|g_{L}\rangle|3,0\rangle+|g_{R}\rangle|0,3\rangle)|000\rangle_{c}|00\rangle_{f},\cr
|\Psi_{4}\rangle&=&\frac{1}{\sqrt{2}}(|g_{L}\rangle|4,0\rangle+|g_{R}\rangle|0,4\rangle)|000\rangle_{c}|00\rangle_{f},\cr
|\Psi_{5}\rangle&=&\frac{1}{\sqrt{2}}(|g_{L}\rangle|5,0\rangle+|g_{R}\rangle|0,5\rangle)|000\rangle_{c}|00\rangle_{f},\cr
&\vdots&\cr
|\Psi_{n}\rangle&=&\frac{1}{\sqrt{2}}(|g_{L}\rangle|n,0\rangle+|g_{R}\rangle|0,n\rangle)|000\rangle_{c}|00\rangle_{f},\cr
&\vdots&
\end{eqnarray}
Here, in the sign $|n,0\rangle(|0,n\rangle)(n=1,2,3,\cdots)$,
$|n\rangle$ donates there are $n$ atoms through the cavity 1(3) in
the state $|k\rangle_{L(R)}$; $|0\rangle$ donates there are $n$
atoms in the state $|s\rangle_{R(L)}$. At last, we use a classical
field $F_{3}$ to implement a Hadamard operation
\begin{eqnarray}\label{20}
|g_{L}\rangle\rightarrow\frac{1}{\sqrt{2}}(|g_{L}\rangle+|g_{R}\rangle),\cr
|g_{R}\rangle\rightarrow\frac{1}{\sqrt{2}}(|g_{L}\rangle-|g_{R}\rangle),
\end{eqnarray}
and then, the state $|\Psi_{n}\rangle$ becomes
\begin{eqnarray}\label{21}
\frac{1}{2}[|g_{L}\rangle(|n,0\rangle+|0,n\rangle)+|g_{R}\rangle(|n,0\rangle-|0,n\rangle)].
\end{eqnarray}
Now, a single projective measurement should be performed on the
double $\Lambda$-type three-level atom. If the double
$\Lambda$-type three-level atom is detected in the state
$|g_{L}\rangle$, Eq. (\ref{21}) collapses to
\begin{eqnarray}\label{22}
|{\rm{NOON}}\rangle_{+}=\frac{1}{\sqrt{2}}(|n,0\rangle+|0,n\rangle),
\end{eqnarray}
and if the atom is detected in the state $|g_{R}\rangle$, Eq.
(\ref{21}) collapses to
\begin{eqnarray}\label{23}
|{\rm{NOON}}\rangle_{-}=\frac{1}{\sqrt{2}}(|n,0\rangle-|0,n\rangle).
\end{eqnarray}
It is worth noting that no matter what the measurement result is,
the NOON states can be always achieved. That's to say, the
successful probability of our protocol is unity in the ideal case.

\section*{4. Discussion and conclusion}

In order to generate the NOON states, the conditions of Eq.
(\ref{11}) and Eq. (\ref{16}) should be satisfied in our scheme.
So we can design the pulse shape of the laser fields $\Omega_{L}$,
$\Omega_{R}$ and $\Omega_{1}$ as the Gaussian
\cite{UG1990,HG2004,YLZ2009},
\begin{eqnarray}\label{25}
\Omega_{\xi}(t)=\Omega_{0}\textmd{exp}[-\frac{(t-T/2-t_{\xi})^2}{2\tau^2}],
\end{eqnarray}
where $\xi=L,R,1;$ $\Omega_{0}$ is the amplitude of
$\Omega_{\xi}$; $T$ is the total adiabatic time and $\tau$ is the
laser beam waist. $t_{\xi}$ is the time we turn on the laser with
Rabi frequency $\Omega_{\xi}$ on the corresponding atoms.
\begin{figure}
\scalebox{0.9}
{\includegraphics[width=3.4in,height=3in]{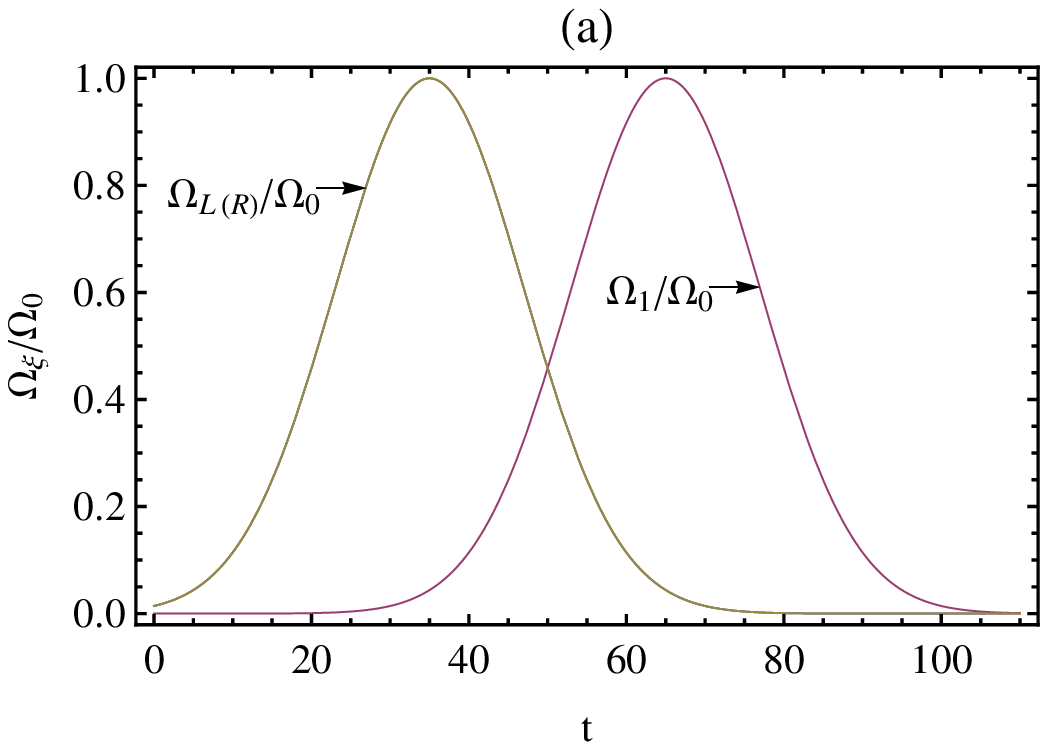}
\includegraphics[width=3.4in,height=3in]{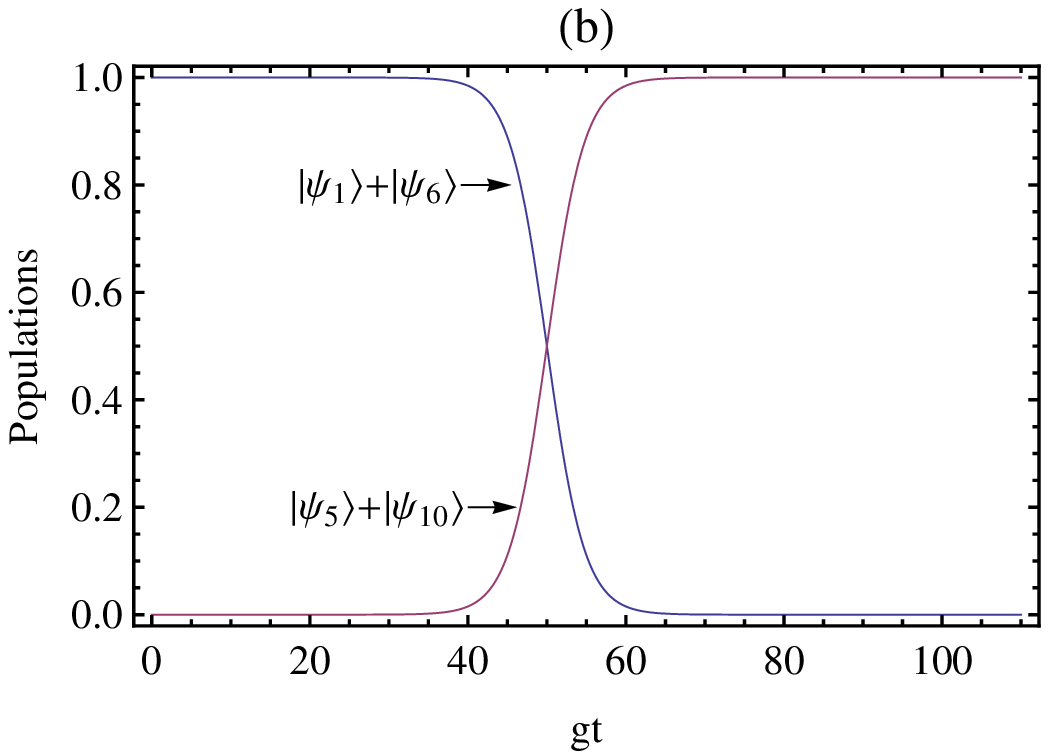}} \caption{(a)
Time dependence of $\Omega_{\xi}(t)/\Omega_{0}$ of the laser
fields for getting $|\Psi_{n}\rangle$. (b) Time evolutions of the
populations of corresponding system states. Here, the system
parameters are set to be $\Omega_{0}= 1.5g, T=100/g, \tau =12/g,
t_{L}= t_{R}= -15/g, t_{1}=15/g, g=1$ GHz and $\Delta =
15g$.}\label{fig04}
\end{figure}

For showing expression (\ref{25}) clearly, we plot the time
dependence of $\Omega_{\xi}(t)/\Omega_{0}$ of the laser fields in
Fig.~\ref{fig04}(a) and the population curves of
$|\psi_{1}\rangle+|\psi_{6}\rangle$ and
$|\psi_{5}\rangle+|\psi_{10}\rangle$ with $gt$ in
Fig.~\ref{fig04}(b). The system parameters are chosen as $g=1$
GHz, $\Omega_{0}=1.5g$, $T=100/g,\tau=12/g, t_{L}=t_{R}=-15/g,
t_{1}=15/g$ and $\Delta=15g$. It can be seen from
Fig.~\ref{fig04}(a) that the conditions
\begin{eqnarray}\label{26}
&&\lim_{t\to-\infty}\frac{\Omega_{1e}}{\Omega_{Le}}=0,~
\lim_{t\to+\infty}\frac{\Omega_{Le}}{\Omega_{1e}}=0,\cr
&&\lim_{t\to-\infty}\frac{\Omega_{1e}}{\Omega_{Re}}=0,~
\lim_{t\to+\infty}\frac{\Omega_{Re}}{\Omega_{1e}}=0,
\end{eqnarray}
can be satisfied during the whole evolution for realizing the NOON
states. In addition, we can see from Fig.~\ref{fig04}(b) that when
$gt\rightarrow+\infty$, the population of
$|\psi_{5}\rangle+|\psi_{10}\rangle$ is 1, which means the NOON
states can be deterministically achieved via adiabatic evolution.

Certainly, all the above results are based on the ideal case
because the influences of atomic spontaneous emission, photon
leakage out of the cavities and fibres are not taken into account.
In fact, the effects of these factors are inevitable, so we will
study the effects of these factors on the fidelity below.

For purpose of generating atomic NOON states, we have used the
large detuning condition and adiabatically eliminated the excited
states of the atoms, as a result, setting the spontaneous emission
rate to be zero is acceptable. While, the effect of photon loss
out of the cavity can also be ignored safely in our scheme because
the cavity modes are never populated in the whole process due to
the adiabatic passage along dark state. The terms related to
$|\psi_{3}\rangle$ and $|\psi_{8}\rangle$, however, have been
discarded for above calculations since we have assumed that
$|\eta_{A}|, |\eta_{B}|\gg |\Omega_{Le}|, |\Omega_{Re}|,
|\Omega_{1e}|$ in Eq. (\ref{10}) and Eq. (\ref{15}). The practical
situation is that the fibre modes may be excited, in other words,
there is a probability that the state
$(|\psi_{3}\rangle+|\psi_{8}\rangle)/\sqrt{2}$ are populated, and
then they may evolve into
$(|g_{L}\rangle|s\rangle_{L1}|s\rangle_{R1}+|g_{R}\rangle|s\rangle_{L1}|s\rangle_{R1})|000\rangle_{c}|00\rangle_{f}/\sqrt{2}$
due to the fibre decays, which will cause error for generating the
NOON states. Considering the effect of photon leakage out of the
fibres, the fidelity can be written as
\begin{eqnarray}\label{24}
F=1-\frac{\gamma_{f}}{2}\int_{0}^{T}\frac{\Omega_{Le}^{2}\Omega_{1e}^{2}}{K_{0}^2}+\frac{\Omega_{Re}^{2}\Omega_{1e}^{2}}{K_{1}^2}dt,
\end{eqnarray}
where $K_{0},K_{1}$ are given in Eq. (\ref{8}) and Eq. (\ref{13});
$\gamma_{f}$ denotes the decay rate of fibres (here we assume the
decay rates of two fibres are the same); $T$ is still the total
adiabatic time.

\begin{figure}
\scalebox{0.9}
{\includegraphics[width=3.4in,height=3.0in]{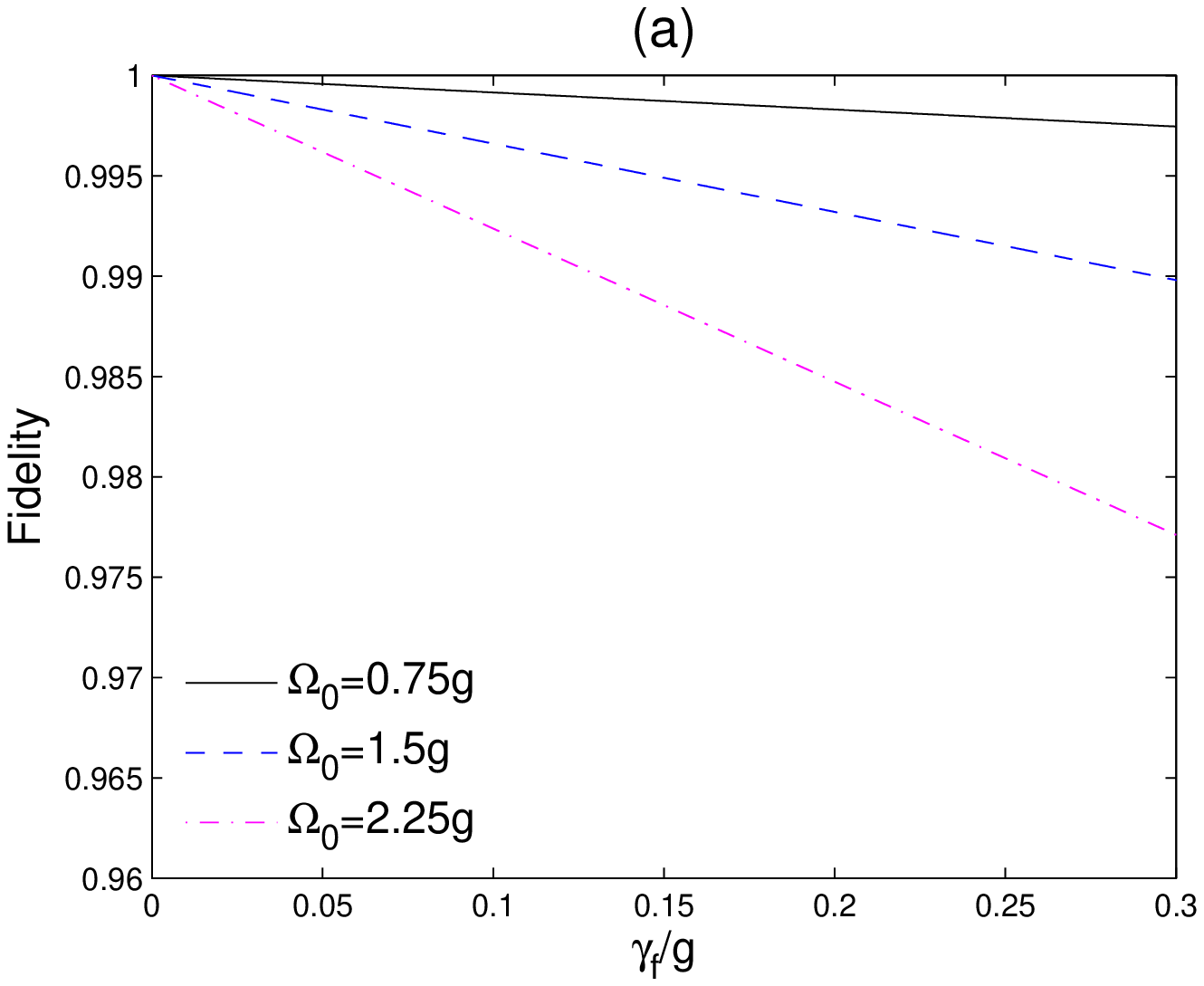}
\includegraphics[width=3.4in,height=3.0in]{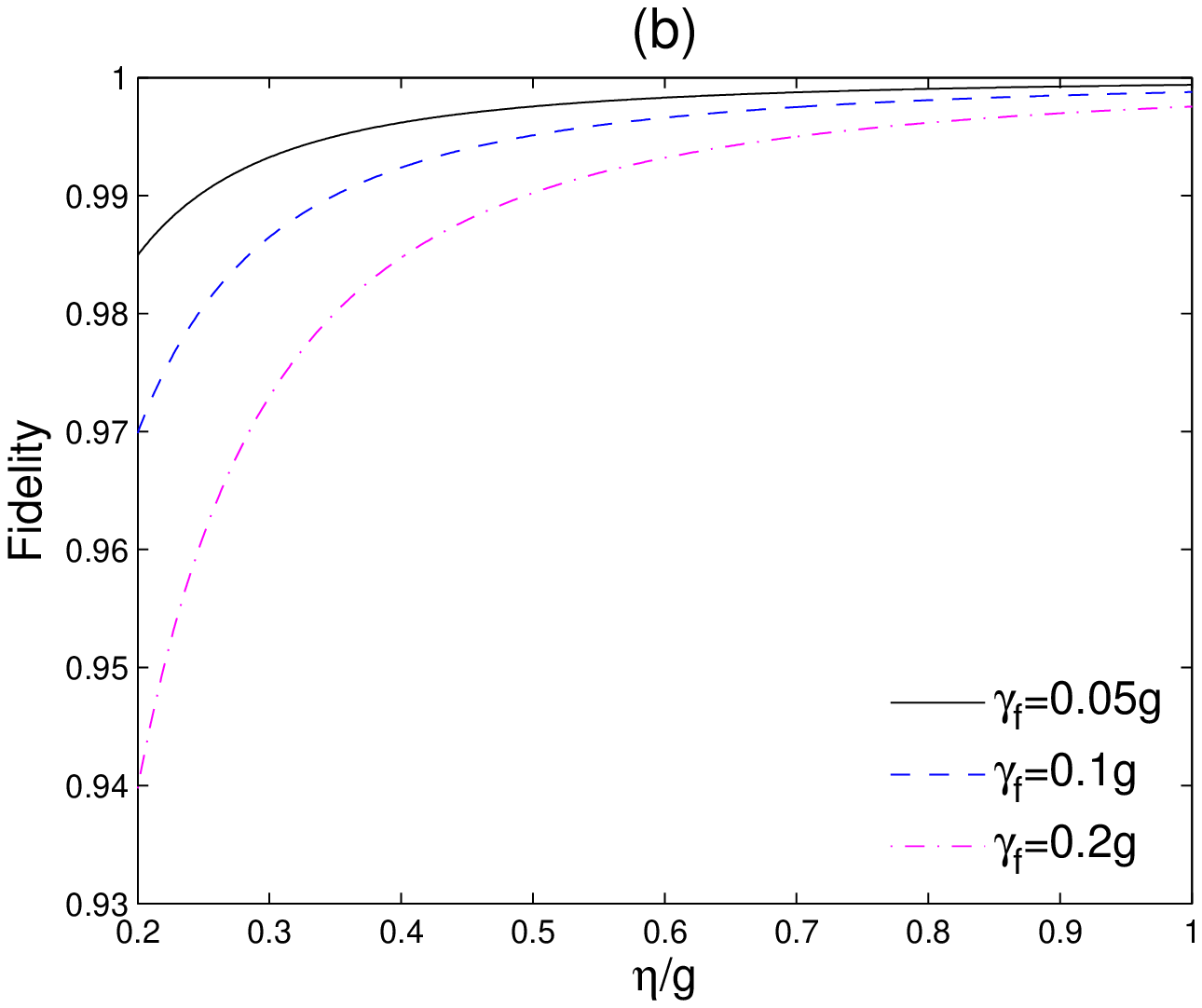}} \caption{The effect of fibre loss $\gamma_{f}$ and the coupling strength
$\eta$ of fibre-cavity on the fidelity of getting
$|\Psi_{1}\rangle$ with $T = 100/g, \tau = 12/g, t_{L} = t_{R} =
-15/g, t_{1} = 15/g, g=1$ GHz and $\Delta = 15g$. (a) Fidelities
each as a function of $\gamma_{f}/g$ for different $\Omega_{0}$
with $\eta=0.6g$. (b) Fidelities each as a function of $\eta/g$
for different $\gamma_{f}$ with $\Omega_{0} = 1.5g$.}\label{fig05}
\end{figure}

We investigate the effect of fibre loss on the fidelity of getting
$|\Psi_{1}\rangle$ for different $\Omega_{0}$ values
($\Omega_{0}=0.75g, 1.5g, 2.25g$) as shown in Fig.~\ref{fig05}(a),
where we set $\eta_{A,B}=\eta=0.6g$. It is seen obviously from
Fig.~\ref{fig05}(a) that the fidelity decreases slightly with the
increase of the $\gamma_{f}/g$. However, even if there exists a
relatively large fibre decay rate $\gamma_{f}=0.3g$ when
$\Omega_{0}\leq1.5g$, we can still obtain NOON states with a high
fidelity. Besides, we plot the fidelity as a function of $\eta/g$
for different $\gamma_{f} (\gamma_{f} = 0.05g, 0.1g, 0.2g)$ in
Fig.~\ref{fig05}(b) with $\Omega_{0}=1.5g$. We can see that the
fidelities increase gradually with the increase of $\eta/g$,
while, with the increase of decay rate $\gamma_{f}$, the fidelity
approaches 1 more slowly. But, when $\eta/g\geq0.6$ the fidelity
is higher than 0.99 even with $\gamma_{f}=0.2g$.
\begin{figure}
\scalebox{0.7}{\includegraphics{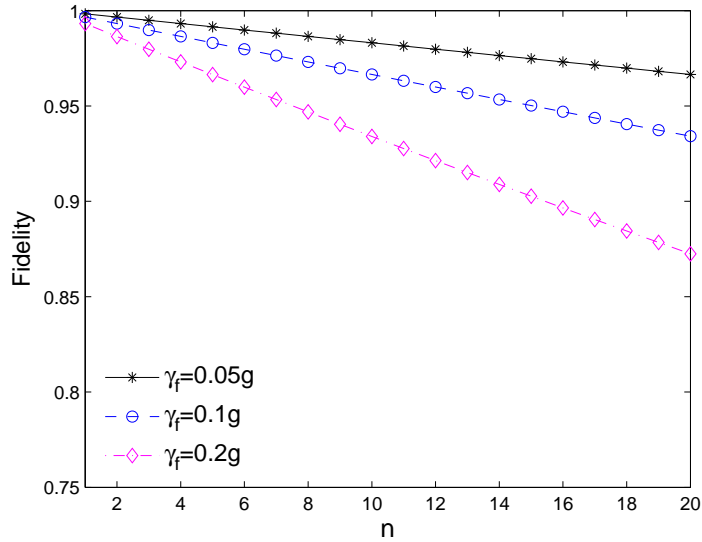}}\caption{Fidelities of
getting $|\Psi_{n}\rangle$ each as a function of $n$ for different
$\gamma_{f}$ when $\Omega_{0}=1.5g$, $\eta=0.6g$ with $T = 100/g,
\tau = 12/g, t_{L} = t_{R} = -15/g, t_{1} = 15/g, g=1$ GHz and
$\Delta = 15g$.}\label{fig06}
\end{figure}

In addition, we also plot the curves of fidelity of getting
$|\Psi_{n}\rangle$ versus $n$ for different decay rates of fibre
modes with $\Omega_{0}=1.5g$ in Fig.~\ref{fig06}. The other
parameters are the same as those in Fig. 5. Obviously, it can be
seen from Fig.~\ref{fig06} that the fidelity decreases with the
increase of the particle number $n$. Nevertheless, the fidelity
can still reach 0.934 even if the particle number $n$ is up to 10
when $\gamma_{f}=0.2g$ and up to 20 when $\gamma_{f}=0.1g$.

In conclusion, we have proposed a scheme for generating arbitrary
large-$n$ NOON states via adiabatic passage. By using a sequence
of pulse laser fields, our atom-cavity-fibre system is always in
the dark states. In the whole process, the influence of atomic
spontaneous emissions, photon leakage out of fibres and cavities
are effectively compressed via adiabatic elimination of excited
states and adiabatic passage. Furthermore, we make an estimation
on the fidelities of the NOON states by considering different
parameters and show that our scheme is insensitive to small
fluctuations of experimental parameters. Anyway, the present
scheme provides an efficient approach to realize arbitrary
large-$n$ atomic NOON states and we hope our work may be useful
for the quantum information in the near future.

\begin{center}$\mathbf{Acknowledgments}$\end{center}
This work was supported by the National Natural Science Foundation
of China under Grant Nos. 11064016 and 61068001.

\end{CJK*}
\end{document}